\documentclass[twocolumn]{aastex631}
\usepackage{graphicx} 
\usepackage{amsmath}
\usepackage{subfigure}
\usepackage[T1]{fontenc}
\usepackage{booktabs}
\usepackage{txfonts}
\usepackage{mhchem}
\usepackage{sistyle}

\newcommand{\rj}{$R_\mathrm{J}$}
\newcommand{\mj}{$M_\mathrm{J}$}

\DeclareMathOperator\erf{erf}

\begin{document}

\title{Impact of Jupiter's heating and self-shadowing on the Jovian circumplanetary disk structure}

\author{Antoine Schneeberger}
\affiliation{Aix- Marseille Universit\'e, CNRS, CNES, Institut Origines, LAM, Marseille, France}

\author{Olivier Mousis}
\affiliation{Aix- Marseille Universit\'e, CNRS, CNES, Institut Origines, LAM, Marseille, France}
\affiliation{Institut Universitaire de France (IUF), France}

\correspondingauthor{Antoine Schneeberger}
\email{antoine.schneeberger@lam.fr}

\begin{abstract}
Deciphering the structure of the circumplanetary disk that surrounded Jupiter at the end of its formation is key to understanding how the Galilean moons formed. Three-dimensional hydrodynamic simulations have shown that this disk was optically thick and significantly heated to very high temperatures due to the intense radiation emitted by the hot, young planet. Analyzing the impact of Jupiter's radiative heating and shadowing on the structure of the circumplanetary disk can provide valuable insights into the conditions that shaped the formation of the Galilean moons.
To assess the impact of Jupiter's radiative heating and shadowing, we have developed a two-dimensional quasi-stationary circumplanetary disk model and used a grey atmosphere radiative transfer method to determine the thermal structure of the disk. We find that the circumplanetary disk self-shadowing has a significant effect, with a temperature drop of approximately 100 K in the shadowed zone compared to the surrounding areas. This shadowed zone, located around 10 Jupiter radii, can act as a cold trap for volatile species such as NH$_3$, CO$_2$ and H$_2$S. The existence of these shadows in Jupiter's circumplanetary disk may have influenced the composition of the building blocks of the Galilean moons, potentially shaping their formation and characteristics. Our study suggests that the thermal structure of Jupiter's circumplanetary disk, particularly the presence of cold traps due to self-shadowing, may have played a crucial role in the formation and composition of the Galilean moons.
\end{abstract}


\section{Introduction}

After Jupiter's formation, the planet is surrounded by a circumplanetary disk (CPD) constituted of gas, dust and ice \citep{lunine1982,canup2002,canup2006,mousis2004,coradini2010,szulagyi2014,szulagyi2016,szulagyi2017,marleau2023,li2023}, in which the Galilean moons are thought to have formed, either in situ \citep{lunine1982,canup2006,chen2020,madeira2021,zhangZhang2021}, or via or through the capture of solar-orbiting planetesimals that later evolved into proto-moons \citep{ronnet2018,ronnet2020,shibaike2019}. 

One category of CPD models postulates that its initial mass might align with the mass of the moon system, supplemented by additional gas to achieve a gas-to-solid ratio comparable to that of the protosolar nebula (PSN) \citep{lunine1982,mosqueira2003}. These CPDs could exhibit masses as substantial as $2 \times 10^{-2}$~\mj, and feature notably warm temperature profiles, reaching several thousand kelvin in the inner regions of the disk \citep{mosqueira2003}. These models are typically very warm, thereby hindering the condensation of water. They likely represent an early phase of the CPD evolution, during which the protoplanet's envelope undergoes significant growth, resulting in a high accretion rate for the CPD, possibly up to $10^{-5}$~\mj/yr \cite{sasaki2010}. In such scenarios, moon formation is presumed to occur through streaming instability and pebble accretion \citep{canup2002,ronnet2018}. However, it remains uncertain whether moons formed in such massive CPDs could have survived, given the potential for type I inward migration induced by satellite-disk interactions, which could be highly efficient \citep{Ward1997,canup2002,alibert2005}. A moon could potentially fall onto the planet over a timescale of approximately 0.1--10 kyr \citep{alibert2005}.

A second category of models proposes a later formation timeline for the moons, occurring when Jupiter's growth was nearing completion. During this phase, Jupiter had carved out a gap in the PSN. Due to the pressure gradient surrounding the gap, only gas and dust situated several scale heights above the midplane could be drawn onto the CPD \citep{lambrechts2012,zhu2012}. Consequently, the net inflow of gas and dust sustaining the subdisk is limited, resulting in the formation of cold CPDs characterized by low accretion rates, typically on the order of $\sim$$10^{-7}$ \mj/yr \cite{canup2002,canup2006,sasaki2010,anderson2021}. These models facilitate the formation of ice at the current orbital locations of the Galilean moons.

However, recent advances in global three-dimensional hydrodynamic simulations point out that runaway gas accretion results in a planet with an exceptionally high surface temperature, reaching several thousand kelvin \citep{szulagyi2016,szulagyi2017}. The incorporation of a third dimension into CPD models has pointed out the existence of meridional circulation patterns, which facilitate the exchange of material between the CPD and the PSN through intricate streams of gas flowing in both directions \citep{tanigawa2012,morbidelli2014,szulagyi2014}. This meridional circulation has profound implications for our comprehension of CPD formation and replenishment processes. By establishing a dynamic physical connection between the PSN and the CPD, it initiates intricate feedback mechanisms that regulate the accretion rate onto the CPD. Consequently, these mechanisms impose constraints on Jupiter's growth during the gas runaway accretion phase \citep{szulagyi2014}. Furthermore, it has been suggested that meridional circulation can uplift dust from the midplane of the PSN, thereby enriching the accreted material with dust--to--gas ratios that surpass those estimated for the PSN \citep{szulagyi2022}. This contrasts with previous arguments proposing significant dust depletion in the accreting material \citep{canup2006,lambrechts2012,zhu2012}.

Three-dimensional simulations also underscore the significance of the opacity regime prescription. In the context of a fully optically thick CPD, studies have demonstrated that its temperature profile can escalate to 4500 K, with a centrifugal radius extending up to one-third of the Hill radius \citep{ayliffe2009}, significantly beyond the extent proposed by the gas-starved model, which posits only a few percent of the Hill radius \citep{canup2002,canup2006}. In the most extreme scenario, \cite{fung2019} and \cite{krapp2024} showed that the gas surrounding Jupiter might be thermally supported, leading to the formation of a circumplanetary shell rather than a CPD. Conversely, when the CPD is assumed to be optically thin and locally isothermal, cooling becomes considerably more efficient, resulting in the formation of a massive CPD \citep{fung2019}.

While three-dimensional simulations offer enhanced accuracy in investigating the interaction between the PSN and the CPD, their computational demands limit their coverage to only a few hundred orbits. In contrast, gas-starved and Minimum Mass Solar Nebula (MMSN) models can span longer timescales on the order of a million years or more. However, these models often overlook the elevated temperature of Jupiter during its formation and the potential presence of an adiabatic regime when the CPD reaches significant mass. Although there are studies that incorporate the planet's internal temperature \citep{makalkin2014,heller2015}, they typically employ a constant opacity prescription and assume a fully optically thick disk, neglecting the resulting disk self-shadowing observed in optically thick protoplanetary disks \citep{dullemond2001,garufi2014,stolker2016,montesinos2021}.

\begin{figure*}
\centering
\includegraphics[trim={1cm 2cm 1cm 1cm},clip, width=1\linewidth]{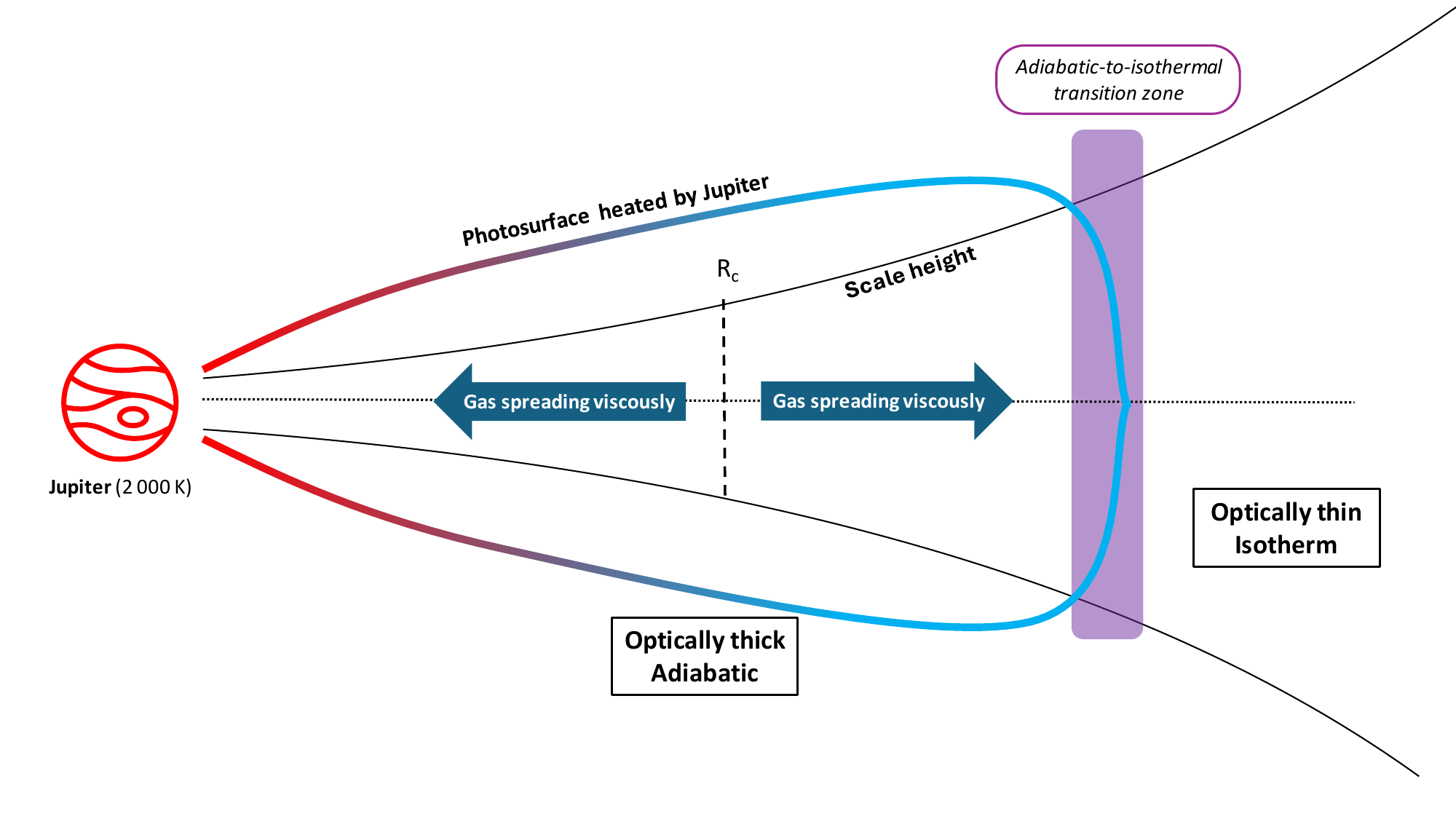}
\caption{Diagram illustrating the structure of the circumplanetary disk model, and delimited into two distinct zones. The first zone, situated closer to Jupiter, is optically thick, characterized by an optical depth $\tau$ exceeding $2/3$ in the midplane. The second zone, located farther from Jupiter, is optically thin, with $\tau$ below $2/3$ in the midplane. The altitude at $\tau = 2/3$ defines the photosurface. Within the optically thick region, bold lines on the diagram depict this photosurface, indicating a temperature gradient ranging from hot (red) to cold (blue) regions. These regions are demarcated by the adiabatic--to--isothermal transition zone (purple box), where the photosurface ceases to exist. Material is uniformly accreted onto the CPD at distances closer to the centrifugal radius ($R_\mathrm{c}$).}
\label{fig:sketch}
\end{figure*}

In this study, our primary objective is to investigate the influence of Jupiter's intrinsic heating and CPD self-shadowing on its structural characteristics. To accomplish this goal, we have developed a two-dimensional CPD model based on the gas-starved disk model and previous research by \cite{makalkin1995}, \cite{makalkin2014}, and \cite{heller2015}. The model architecture, illustrated in Figure \ref{fig:sketch}, divides the CPD into two distinct regions: an adiabatic, optically thick region and an isothermal, optically thin region. The heating of the CPD by Jupiter's radiative heating occurs at its photosurface, where the medium becomes optically thick. This heat is subsequently propagated to the midplane using a grey atmosphere radiative transfer model. Furthermore, we account for the projection of shadows cast by the optically thick region onto the optically thin region. Our investigation encompasses a wide range of CPD metallicities, spanning from subsolar values \citep{lambrechts2012,zhu2012} to supersolar values \citep{szulagyi2022}.

Section \ref{sec:method} is devoted to describing the two-dimensional model of the circumplanetary disk (CPD). Section \ref{sec:results} presents the findings from simulations employing various metallicity prescriptions, along with results concerning the self-shadowing of the CPD. Discussion on the implications and limitations of our approach, as well as an examination of parameter sensitivity, is provided in Section \ref{sec:discussion}. Finally, Section \ref{sec:summary} summarizes the study and presents the conclusions.

\section{Model}
\label{sec:method}

In this section, we provide a detailed description of the two-dimensional CPD model employed in our simulations, including the time evolution and self-shadowing prescriptions incorporated into the model.

\subsection{Density profile of the CPD}
\label{sec:density_profile}

The CPD is axisymetric and at hydrostatic equilibrium. We adopt the viscous accretion model  of  \citep{canup2002,canup2006} , where the viscosity is parametrized by the $\alpha$--turbulence viscosity of \cite{shakura1973} .  In this model, the CPD is fed by an uniform flux of gas and dust in centrifugally supported orbits, located at distances shorter than the so-called centrifugal radius $R_c$. At distance larger than the centrifugal radius, infalling material is ejected back to the PSN and does not accrete onto the CPD.  The surface density is obtained by solving the continuity equation for the infalling gas  \citep{canup2002}:

\begin{equation}
\Sigma_\mathrm{g} = \frac{\dot{M}}{3 \pi \nu} \frac{\Lambda}{l},
\label{eq:surface_density}
\end{equation}

\noindent where $\dot{M}$ is the accretion rate of matter onto the planet, and $\nu$ is the gas viscosity defined as 

\begin{equation}
\nu = \alpha \frac{C_s}{\Omega_\mathrm{K}}.
\label{eq:viscosity}
\end{equation}

\noindent  Here, $C_s$ denotes the sound speed in the gas, $\Omega_\mathrm{K}$ represents the Keplerian pulsation, and $\alpha$ is a free parameter whose value can vary within the range of $10^{-5}$ to $10^{-2}$, as indicated by disk observations \citep{shakura1973,lynden-bell1974,villenave2022}. The sound speed $C_s$ is defined as follows: 

\begin{equation}
C_s = \sqrt{\frac{RT}{\mu_g}},
\label{eq:sound_speed}
\end{equation}

\noindent with $R$ representing the ideal gas constant, $T$ denoting the gas temperature, and $\mu_g$ representing the gas mean molar mass (2.31 g·mol$^{-1}$). Additionally, the factors $\Lambda$ and $l$ are derived from the gas surface density continuity equation \citep{canup2002,makalkin2014}. When considering distances shorter than $R_c$, the following relation holds:

\begin{equation}
\begin{split} 
\Lambda(r) =  &1 - \frac{1}{5} \left( \frac{r}{R_c} \right)^2 - \sqrt{\frac{R_p}{r}}- \frac{4}{5} \sqrt{\frac{R_c}{r_d}} \\ 
&+ \frac{4}{5} \frac{R_p R_c}{r_d r} + \frac{1}{5}\sqrt{\frac{R_p}{r_d}} \left( \frac{r}{R_c} \right)^2,
\end{split}
\end{equation}

\noindent while at distance greater  than $R_c$ 

\begin{equation}
\begin{split}
\Lambda(r) =  \frac{4}{5} \sqrt{\frac{R_c}{r}} - \frac{R_p}{r} - \frac{4}{5} \frac{R_c}{r_d} + \sqrt{\frac{R_p}{r_d}},
\end{split}
\end{equation}

\noindent with $R_p$ representing the radius of the planet. Finally one have: 

\begin{equation}
l = 1 - \sqrt{\frac{R_p}{r_d}},
\end{equation}

\noindent with $r_d$ denoting the size of the disk.

At the centrifugal radius location, the centrifugal force of an object with a specific angular momentum compensates for the gravitational force. Closer than this radius, the CPD's gas will drift inward toward the planet, while at greater distances, it flows outward, merging with the meridional flow within the Hill sphere of the planet \citep{szulagyi2016,szulagyi2017,batygin2020}. Therefore, the centrifugal distance is defined as:

\begin{equation}
R_c = \frac{j^2}{GM_p},
\label{eq:centrifugal_radius}
\end{equation}

\noindent where $j$ represents the specific angular momentum of infalling gas from the Hill sphere, $G$ denotes the universal gravitational constant, and $M_p$ stands for the mass of the planet. The centrifugal radius exhibits significant variability, contingent upon the value of the specific angular momentum parameter $j$. The centrifugal radius can then be treated as an input parameter in the model, allowing for the exploration of different scenarios and configurations. Since the moon's building blocks are hypothesized to have formed within the centrifugal radius, this radius must extend beyond the orbit of the outermost moon, Callisto, located at 26.9~\rj \citep{canup2006,sasaki2010,heller2015}.

\subsection{Vertical structure}
\label{sec:vert_struct}

Assuming a CPD at hydrostatic equilibrium, the vertical density profile follows a Gaussian distribution:

\begin{equation}
\rho(r,z) = \rho_0(r) \exp{\left(-\frac{z^2}{2h(r)^2}\right)},
\label{eq:rho(r,z)}
\end{equation}

\noindent where $\rho_0$ represents the gas density at the disk midplane. $h$ denotes the disk scale height, given by:

\begin{equation}
h = \frac{C_s}{\Omega_\mathrm{K}}.
\label{eq:scale_height}
\end{equation}

By definition, the surface density $\Sigma_g(r)$ is given as the integral of the density $\rho(z,r)$ over all vertical heights:

\begin{equation}
\Sigma_g(r) = \int_{-\infty}^{+\infty} \rho(z,r) dz.
\label{eq:surface_density_formula}
\end{equation}

\noindent one can thus deduce the midplane density by combining this equation with the vertical density profile:

\begin{equation}
\rho_0(r) = \sqrt{\frac{2}{\pi}} \frac{\Sigma_g}{2h},
\label{eq:midplane_density}
\end{equation}

\noindent which depends solely on the distance to the planet.

The CPD is divided into two regions: an adiabatic, optically thick zone where the optical depth $\tau$ in the midplane exceeds the standard value of $2/3$, and a vertically isothermal, optically thin zone where $\tau$ is less than $2/3$ in the midplane (see Fig. \ref{fig:sketch}). $\tau$ is defined as:

\begin{equation}
\tau(z) = \int_z^{+\infty} \rho(r,z') \kappa_\mathrm{R}(r,z') dz',
\label{eq:opt_depth}
\end{equation}

\noindent where $\kappa_\mathrm{R}$ represents the mean Rosseland opacity of the medium.In this study, we use the opacity experimental data compiled by \cite{pollack1994} and fitted by \cite{makalkin2006} into a piecewise continuous power function :

\begin{equation}
    \kappa_\mathrm{R} = \chi \kappa_0 T^{\beta} 
\end{equation}
\noindent where $\chi$ is the dust enrichment factor between the protoplanetary disk and the CPD, $\kappa_0$ and $\beta$ are opacity parameters fitted from the experimental data. The parameters and opacity regimes are detailed in table \ref{table:opacity}.

\begin{table*}
\caption{Opacity regimes parameters, adapted from \cite{pollack1994} and \cite{makalkin2006}}
\label{table:settings}
\centering
    \begin{tabular}{cccc}
    \hline\hline    Regime & Temperature (K) range & $\beta$ & $\kappa_0$ (cm$^{2}$.g$^{-1}$) \\
    \hline
     Water ice & T < 173 & 2.1 & $1.6 \times 10^{-4}$ \\
     Vaporization of water & 173 < T < 425 & 0.6 & 0.17 \\
     Vaporization of organic matter & 425 < T < 680 & 0.5 & 0.1 \\
     Vaporization of refractory minerals & T > 680 & 0.75 & $1.9 \times 10^{-2}$\\ 
    \hline
    \end{tabular}
    \label{table:opacity}
\end{table*}

The optically thick region is heated by the planet's radiation at the CPD photosurface level, corresponding to $\tau$ = $2/3$. By combining the density profile equation with the definition of optical depth, one can determine the altitude of the photosurface, $z_s$, by solving:

\begin{equation}
\frac{2}{3} = \int_{z_s}^\infty \rho(r,z') \kappa_\mathrm{R}(r,z') dz'.
\label{eq:photo_opt_depth}
\end{equation}

Since the region above the photosurface is vertically isothermal, the opacity is constant above the photosurface. This allows for computing $z_s$ by combining Eqs. \ref{eq:rho(r,z)},\ref{eq:midplane_density},\ref{eq:photo_opt_depth}:

\begin{equation}
z_s(r) = \mathrm{erf}^{-1}\left( 1 - \frac{4}{3} \frac{1}{\Sigma_g(r) \kappa_R(r,z_s)} \right) \sqrt{2} h(r),
\end{equation}

\noindent where $\kappa_p(r,z_s)$ denotes the opacity at the photospheric location. 

The photosurface exists only if $z_s$ is positive, leading to the condition:

\begin{equation}
\begin{split}
&\erf^{-1} \left(1 - \frac{4}{3} \frac{1}{ \Sigma_g(r) \kappa_p(r,z_s)}\right) > 0 \\ 
& \iff  \Sigma_g(r) \kappa_p(r,z_s) > \frac{4}{3}.
\end{split}
\end{equation}

From this condition, the radial position of the transition zone between an optically thick CPD and a fully optically thin CPD can be determined. The CPD is optically thin if:

\begin{equation}
\Sigma_g(r) \kappa_p(r,0) < \frac{4}{3},
\end{equation}

\noindent where $\kappa_p(r,0)$ represents the opacity at the CPD midplane location.

\subsection{Thermal structure in the optically thick region} 
\label{sec:thick_region}

In the optically thick region, the temperature at the photosurface altitude, referred to as the surface temperature, follows the semi-analytical model \citep{makalkin1995,makalkin2014}:

\begin{equation}
\begin{split}
T_s^4(r) = & \frac{1 + \left( 2 \kappa_p(r,z_s) \Sigma_g(r) \right)^{-1}}{\sigma_{\mathrm{SB}}}\\
&\times \left( F_{\mathrm{vis}} +  F_{\mathrm{acc}} + k_s F_{\mathrm{p}} \right) + T_{\mathrm{neb}}^4.
\end{split}
\label{eq:surface_temperature}
\end{equation}

\noindent where $\sigma_{\mathrm{SB}}$ represents the Stefan-Boltzmann constant, $T_{\mathrm{neb}}^4$ denotes the temperature of the surrounding PSN, and $F_{\mathrm{vis}}$, $F_{\mathrm{acc}}$, and $F_{\mathrm{p}}$ are the heat fluxes produced by viscous stress, accretion of material onto the CPD, and Jupiter's radiative heating, respectively. Additionally, $k_s$ is a factor between 0 and 1 defining the fraction of Jupiter's light absorbed, with the remainder either reflected or scattered. We use the value $k_s = 0.2$ \citep{makalkin1995}. The expressions for the heat fluxes are then \citep{makalkin2014}:

\begin{equation}
F_{\mathrm{vis}} = \frac{3}{8 \pi} \frac{\Lambda}{l} \dot{M} \Omega_\mathrm{K}^2,
\end{equation}

\begin{equation}
F_{\mathrm{acc}} = \frac{X_d \chi G M_p \dot{M}}{4 \pi R_c^2r}\exp{\left(-\frac{r^2}{R_c^2}\right)},
\end{equation}

\begin{equation}
F_{\mathrm{p}} = L_\mathrm{p} \frac{ \sin{(\zeta + \eta)} }{8 \pi (r^2 + z_s^2)}.
\end{equation}

\noindent Here, $X_d$ represents the metallicity of the PSN, and $\chi$ is the metallicity enrichment factor between the PSN and the CPD, associated with dust enrichment. In the expression for Jupiter's surface heating flux, $L_p$ denotes the planet's luminosity, while $\zeta$ and $\eta$ are angles expressing the disk's geometry, defined as follows:

\begin{equation}
\zeta = \arctan{ \left( \frac{4}{3 \pi} \frac{R_p}{\sqrt{r^2 + z_s^2}} \right)},
\end{equation}

\begin{equation}
\eta = \arctan{ \left( \frac{d z_s}{d r} \right) } - \arctan{ \left( \frac{z_s}{r} \right)}.
\end{equation}

From the surface temperature, we determine the vertical temperature structure using the Eddington approximation \citep{makalkin1995}. By utilizing the first and second moments of the radiative transfer equation and considering that Jupiter's light does not penetrate the optically thick region, we arrive at:

\begin{equation}
\frac{d T^4}{dq} = -\frac{27}{64 \pi} \nu(q) \Sigma_g^2 \Omega_\mathrm{K}^2 \kappa_\mathrm{R}(q) q,
\label{eq: rad_trans}
\end{equation}

\noindent where $q$ represents the vertical mass coordinate given by:

\begin{equation}
q(z) = 2 \int_0^z \rho(z') dz'.
\end{equation}

Normalizing the mass coordinate in Eq. \ref{eq: rad_trans} by the mass coordinate of the photosurface $q_s$, we obtain:

\begin{equation}
\frac{d T^4}{dq'} = -\frac{27}{64 \pi} \nu(q') \Sigma_g^2 \Omega_\mathrm{K}^2 \kappa_\mathrm{R}(q') q' q_s^2,
\label{eq: rad_trans_normalized}
\end{equation}

\noindent where

\begin{equation}
q' = \frac{q}{q_s}.
\end{equation}

\noindent Here, the normalized mass coordinate $q'= 1$ corresponds to the altitude of the photosurface, implying:

\begin{equation}
T(q' = 1) = T_s.
\end{equation}

\noindent We assume that the disk is vertically symmetric with respect to the midplane, implying that the vertical temperature profile attains a maximum value at the midplane altitude:

\begin{equation}
\left. \frac{dT}{dq'} \right|_{q'=0} = 0.
\end{equation}
 
 In the aforementioned equation system, the viscosity $\nu$, the surface density $\Sigma_g$, and the mean Rosseland opacity $\kappa_\mathrm{R}$ depend on the temperature itself. Furthermore, the surface temperature depends on the derivative of the photosurface altitude, which itself depends on the midplane temperature. We employ an iterative solving method based on the Jacobian-Free-Newton-Krylov method (JFNK) \citep{Knoll1993,Knoll2004,Zhou2022} to solve Eq. \ref{eq: rad_trans_normalized} in the $(r,q')$ parameter space. Subsequently, we use a spline cubic interpolation method to transform the solution from $(r,q')$ space to $(r,z)$ space. 

\subsection{Temperature in the optically thin region}
\label{sec:thin_region}

When the disk is vertically isothermal, the disk surface temperature given by Eq. \ref{eq:surface_temperature} becomes its effective temperature. However, in the absence of a well-defined photosurface, a smaller fraction of Jupiter's incident radiation is absorbed by the medium within the CPD. Hence, we define an absorption factor based on the midplane optical depth:

\begin{equation}
\tau(r,z=0) = \int_0^\infty\rho(r,z')  \kappa_R(r,z')  dz'.
\end{equation}
Given that we are operating in the optically thin region, and assuming a vertically isothermal temperature profile, $\kappa_R$ remains constant with respect to the vertical coordinate, $z'$. Thus, we have:

\begin{equation}
\tau(r,z=0) = \frac{\kappa_R \Sigma_g}{2}. 
\end{equation}

As a result, only a fraction of Jupiter's light is absorbed by the disk. We employ a simple Beer-Lambert law approximation:

\begin{equation}
F_{\rm a} = F_{\rm p} e^{-\tau},
\end{equation}

\noindent where $F_a$ represents the absorbed flux. The effective temperature is then computed using Eq. \ref{eq:surface_temperature}, with the planetary flux $F_{\rm p}$ replaced by the absorbed flux $F_{\rm a}$.

\subsection{Self-shadowing}
\label{sec:shadow}

The photosurface, being completely opaque, can cast a shadow onto part of the CPD, either onto the photosurface itself or onto the optically thin area. Particularly at the frontier between the optically thick and optically thin regions, the photospheric altitude decreases, forming an opaque wall in this region. Materials close to the wall are then plunged into its shadow. When a part of the disk is in shadow, it is no longer heated by Jupiter's light, resulting in significantly colder temperatures compared to its surroundings.

A part of the disk is considered to be shadowed if it fulfills the following condition:

\begin{equation}
z_s(r) > R_\mathrm{J} + \frac{z_s(r' - R_\mathrm{J})}{r'}   r,  \quad \forall r' < r.
\end{equation}

\noindent Here, $z_s(r)$ is equal to 0 when the model is operating in an optically thin region.

\subsection{Time evolution}
\label{sec:evol}

The model operates as a quasi-static framework, assuming hydrostatic equilibrium at each time step. Consequently, the properties of the CPD, including its mass and thermodynamic characteristics, depend solely on the time-varying CPD accretion rate and the planet's luminosity. We adopt a Jupiter accretion model from \cite{mordasini2012a,mordasini2013}, which model the accretion process by considering a known PPD density profile and gap formation around Jupiter. From this model, we derive a simple parametric fit:

\begin{equation}
\dot{M}(t) = \exp\left(- 61.5 \times \frac{t}{10^{6} \mathrm{yr}} - 12.0\right) M_{\mathrm{J}}/\mathrm{yr}
\end{equation}

\noindent and

\begin{equation}
L_p(t) = \frac{1}{79285.7 \frac{t}{10^6 \mathrm{yr}} - 12.442 } L_{\odot},
\end{equation}

\noindent where $t$ is the time elapsed since the end of the runaway gas accretion, which is assumed to be the time of formation of the CPD.  At each time step, the thermodynamic properties of the CPD are computed based on the corresponding accretion rate and luminosity.

\section{Results} 
\label{sec:results}

Simulations were conducted spanning a time range from 30 kyr to 400 kyr following the formation of the CPD. We first present the results for the nominal case, where the CPD metallicity is set to 0.1 times the protosolar value. This subsolar metallicity factor is motivated by the observations from hydrodynamic models, which suggest that the gas feeding the CPD has a subsolar metal content \citep{canup2002,lambrechts2012,zhu2012}. We then compare these results with cases of larger metallicities, 1 and 10 times the protosolar value, to explore scenarios where dust pileup occurs in the CPD, as proposed by \cite{drazkowska2018} and \cite{batygin2020}.

We assume that Jupiter's formation occurred 1 Myr after the formation of calcium-aluminum-rich inclusions (CAIs) \citep{mordasini2013}. In our study, we do not account for Jupiter's post-formation migration and keep the planet at its present-day distance of 5 AU from the Sun. We assume a centrifugal radius of 50~\rj, significantly beyond Callisto's current orbit, to account for the potential inward migration of the moons following their formation \citep{shibaike2019}.

Finally, we set the minimum temperature of the CPD to 40 K, which corresponds to the surrounding PSN temperature after 1 Myr following CAI formation, according to \cite{aguichine2020}. Detailed specifications of all model parameters are provided in Table \ref{table:init_run}.  

\begin{table}[h]
\caption{Nominal model parameters}
\label{table:settings}
\centering
    \begin{tabular}{l ll}
    \hline\hline    Parameter& Value  &References\\
    \hline
    M$_\mathrm{p}$ & 0.95 \mj  & \cite{mordasini2013}\\
    R$_\mathrm{p}$ & 1.1 \rj  & \cite{mordasini2013}\\
    a$_\mathrm{p}$ & 5 AU  & \\ 
    t$_\mathrm{initial}$ & 3 $\times 10^{4}$ yr &\\
    t$_\mathrm{final}$  & 4 $\times 10^{5}$ yr &\\
    X$_d$ & 2.45 $\times 10^{-2}$ & \cite{Lodders2019}\\
    $\chi$ & $10^{-1}$ --  10 &\\
    $k_s$ & $0.2$  & \cite{makalkin1995}\\
    $\alpha$ & $10^{-3}$  & \cite{lynden-bell1974}\\
    $T_\mathrm{amb}$ & 40 K  & \cite{aguichine2020}\\
    $\mu_g$ & $2.341 \times 10^{-3}$  & \cite{Lodders2019}\\
    $R_c$ & 50 \rj  &\\
    \hline
    \end{tabular}
    \label{table:init_run}
\end{table}

\subsection{Nominal model results}

\begin{figure}[h]
\centering
\includegraphics[width=1\linewidth]{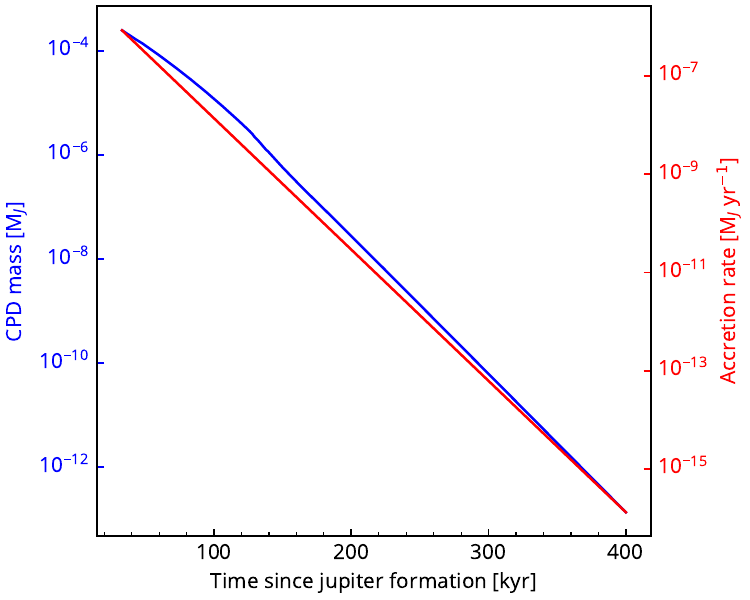}
\caption{Time evolution of the CPD's mass (blue line) and gas accretion rate (red line).}
\label{fig:macc-mass}
\end{figure}

Figure \ref{fig:macc-mass} illustrates the temporal evolution of both the CPD mass and Jupiter's accretion rate over a time span of 30 kyr to 400 kyr following the formation of the CPD. The figure vividly portrays the rapid depletion of the CPD, with the accretion rate ranging from $8.4\times 10^{-7}$ M$_\mathrm{J}$.yr$^{-1}$ to $1.3\times 10^{-16}$ M$_\mathrm{J}$.yr$^{-1}$, encompassing a remarkable nine orders of magnitude. Concurrently, the CPD mass undergoes a dramatic reduction from an initial value of $1.26\times 10^{-4}$ M$_\mathrm{J}$ to a final value of $6.46\times 10^{-11}$ M$_\mathrm{J}$, spanning seven orders of magnitude.

\begin{figure}
\centering
\includegraphics[width=\columnwidth]{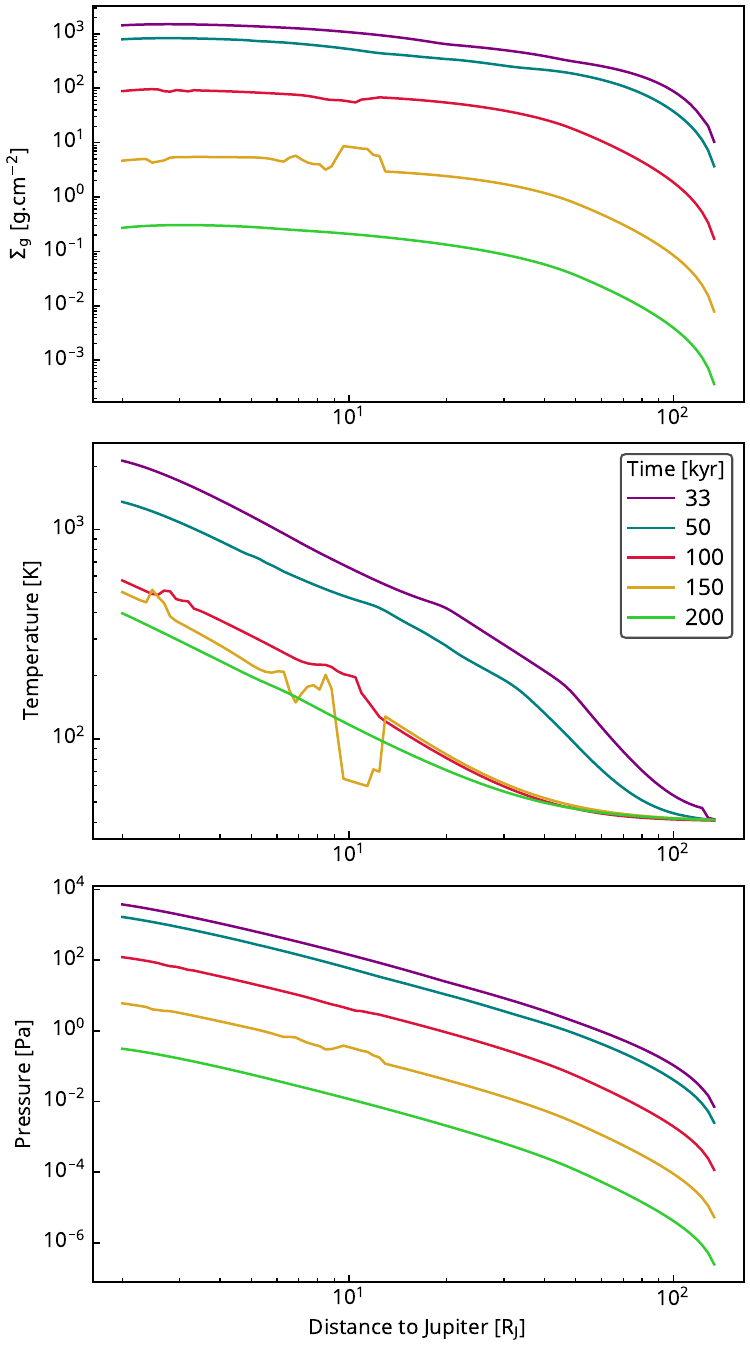}
\caption{From top to bottom: CPD's surface density, midplane temperature and midplane pressure profiles, represented at $t$ = 33, 50, 100, 150 and 200 kyr of the CPD evolution.}
\label{fig:PTSig}
\end{figure}

Figure \ref{fig:PTSig} presents the surface density, midplane temperature, and midplane pressure profiles of the CPD at 33, 50, 100, 150, and 200 kyr after CPD formation. It showcases the rapid depletion of the CPD, with the surface density diminishing from a range between $1.3 \times 10^3$ and $10^2$ g.cm$^{-2}$ 33 kyr, to a range between $2 \times10^{-1}$ and $3.6 \times10^{-4}$ g.cm$^{-2}$ 200 kyr, after Jupiter's formation. Concurrently, the CPD's temperature experiences a decline from a peak of 2270 K at 33 kyr to approximately 440 K at 200 kyr. Consequently, over the same time span, the midplane pressure diminishes by a factor of $10^4$, consistent with the rapid CPD depletion depicted in Figure \ref{fig:macc-mass}.

Before 100 kyr, the primary sources of heating in the CPD are the viscous and accretion processes. However, following 100 kyr, as a result of the rapid depletion of the CPD, radiative heating from Jupiter's hot surface becomes predominant. This transition is explored in Sec. \ref{sec:thick_region}, where the dependence of the disk's heating on the slope of the photosurface, due  to Jupiter's illumination, is emphasized. Figure \ref{fig:PTSig} further illustrates that beyond 100 kyr, the temperature distribution does not exhibit a monotonic decrease with distance from Jupiter. At around 150 kyr post-Jupiter's formation, a narrow region centered around 2.5 \rj~exhibits a temperature 50 K higher than its surroundings, persisting for approximately 10 kyr. Conversely, a region between 9 and 15 \rj appears to be 100 K colder than its surroundings, lasting for about 60 kyr, spanning from 100 kyr to 160 kyr after the CPD's formation.

\begin{figure*}[h]
\centering
\includegraphics[width=\linewidth]{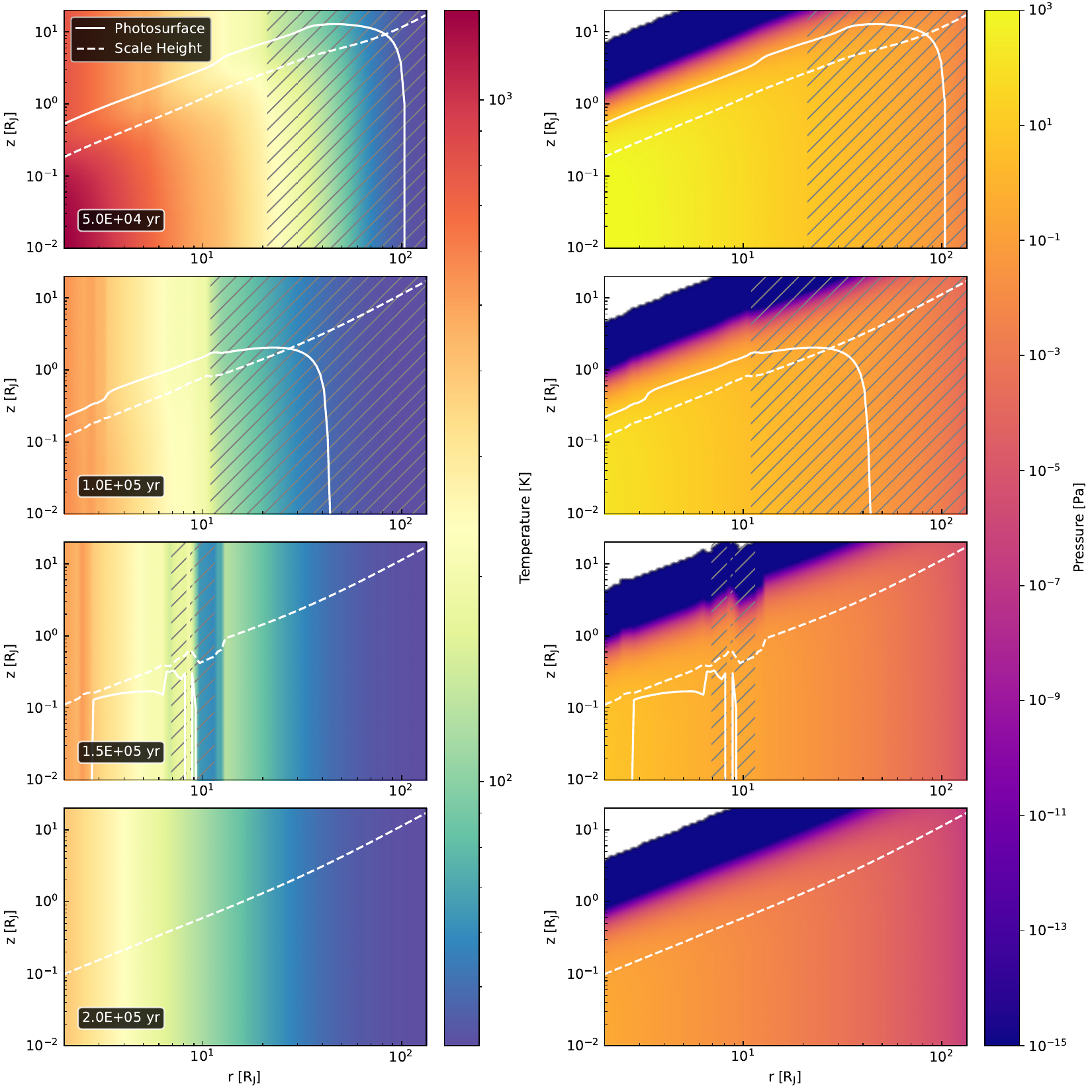}
\caption{Temperature (left panels) and pressure (right panels) profiles of the CPD at $t$ = 50, 100, 150, and 200 kyr of the CPD evolution. The photosurface and the CPD's scale height are presented by the white plain and dashed lines, respectively. . If the photosurface is not explicitly shown or represented in a given panel, it indicates its absence. The hatched area is located within the shadow cast by the CPD. }
\label{fig:temperature_profiles}
\end{figure*}

 Figure \ref{fig:temperature_profiles} provides a visual representation of the two-dimensional temperature and pressure profiles of the CPD at $t$ = 50, 100, 150, and 200 kyr of its evolution. At 100 kyr of CPD evolution, the temperature difference between the photosurface and the midplane remains small, typically within 1 K. This suggests that the disk is nearly vertically isothermal, emphasizing the dominance of Jupiter's radiative heating at the photosurface altitude over viscous and accretion heating mechanisms. At 150 kyr of CPD evolution, an optically thin region forms in the inner part of the CPD, specifically in regions closer than 3 \rj. In these regions, where the temperature exceeds 700 K, the partial evaporation of the most abundant dust components, such as Mg-silicates and iron, leads to a sudden drop in opacity from the experimental data used in our model. This phenomenon induces a decrease in the photosurface altitude at the evaporation front, eventually causing the photosurface to vanish in the inner part of the disk when the density reaches sufficiently low levels. The photosurface has entirely vanished after 200 kyr of CPD evolution, resulting in the partial absorption of Jupiter's radiation. Consequently, in the absence of shadows causing cold regions or flaring inducing hot regions, the temperature profile exhibits a monotonic decrease with distance from Jupiter.

\subsection{Disk metalllicity}

\begin{figure*}[h]
\centering
\includegraphics[width=\linewidth]{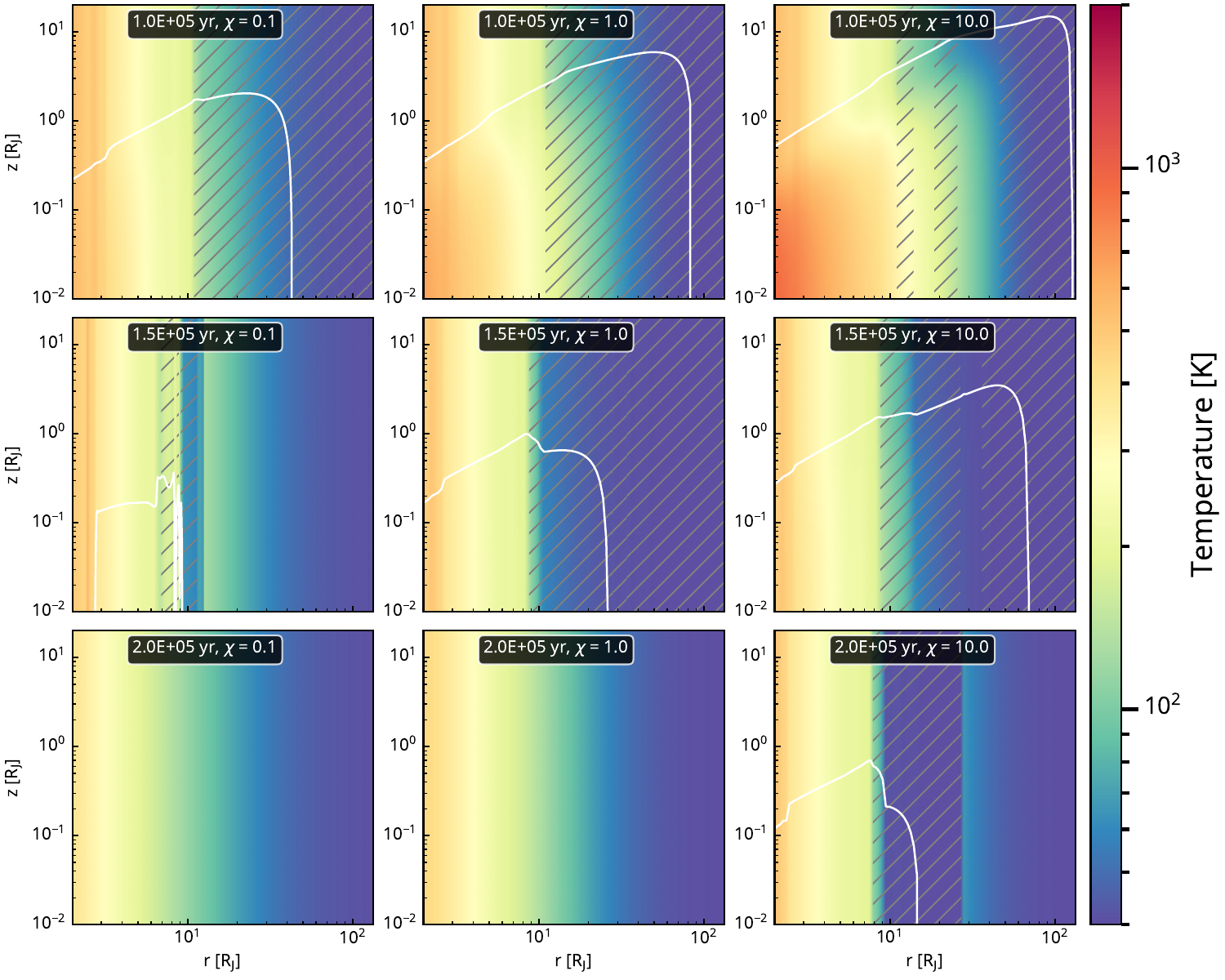}
\caption{CPD temperature profiles with dust enrichment levels (from left to right columns) corresponding to 0.1, 1, and 10 times the protosolar value, depicted at $t$ = 100, 150, and 200 kyr of the CPD evolution (from top to bottom rows). The solid white line represents the altitude of the photosurface. If the photosurface is not explicitly shown or represented in a given panel, it indicates its absence. The hatched areas are located within the shadow cast by the CPD.}
\label{fig:time-metalicity}
\end{figure*}

Figure \ref{fig:time-metalicity} represents three snapshots of the CPD at $t$ = 100, 150, and 200 kyr of its evolution. The three simulations represented in the figure differ in their metallicity enrichment levels, corresponding to 0.1, 1, and 10 times the protosolar metallicity ($Z/H = 2.45 \times 10^{-2}$; \cite{Lodders2019}). The figure demonstrates that an increase in metallicity corresponds to higher opacity within the CPD, consequently leading to an enlarged adiabatic zone relative to cases with lower metallicity levels. In the nominal simulation, with a metallicity 0.1 times protosolar, the adiabatic–isothermal transition zone is located at 46.5 \rj~and 12.9 \rj~after 100 and 150 kyr of CPD evolution, respectively. However, this transition disappears after 200 kyr. For simulations with protosolar metallicity, the adiabatic–isothermal transition occurs at distances of 88.2 \rj~and 34.5 \rj~after 100 and 150 kyr of CPD evolution, respectively. In the scenario where the CPD exhibits a metallicity ten times higher than the protosolar value, this transition is observed at distances of 127.7 \rj, 71.9 \rj, and 25.2 \rj~after 100, 150, and 200 kyr of CPD evolution, respectively. Furthermore, the adiabatic zone vanishes after 161 kyr, 197 kyr and 230 kyr following the CPD formation, for a metallicity of 0.1, 1 and 10 times the protosolar value respectively.

The increased opacity leads to two primary consequences. First, it results in an elevated photosurface altitude, allowing it to cast shadows further into the CPD. For instance, after 150 kyr of CPD evolution, a shadow extends from 9 to 15 \rj~in the nominal simulation, with a metalicity 0.1 times the protosolar value, disappearing after 161 kyr of CPD evolution.  In contrast, for one and tenfold enrichments, shadows extend from 10 to 133 \rj~and from 9.2\rj~to 133 \rj, respectively, after 150 kyr of CPD evolution. In the scenario with a tenfold enrichment in metallicity, the photosurface exhibits a remarkably prolonged existence compared to the other cases. It casts a shadow from 9.5 $R_\mathrm{J}$ to the outer edge of the disk until 196 kyr of CPD evolution, persisting until it ultimately dissipates after 230 kyr. Second, the augmented opacity results in a substantial escalation of temperatures. After 100 kyr of CPD evolution, the maximum midplane temperatures attain values of 700, 720, and 1009 K for simulations with metallicity enrichments of 0.1, 1, and 10 times the protosolar value, respectively.

\subsection{$\alpha$-viscosity factor}
\label{sec:alpha_visco}

The source of viscosity in the PPD and the CPD remains open question. In our model, we adopt the $\alpha$-viscosity prescription, setting the $\alpha$ factor to the conventional value of $10^{-3}$ \citep{shakura1973,lynden-bell1974}. However, it's worth noting that the specific value of $\alpha$ can vary significantly depending on the underlying physical mechanisms considered. For a fully ionized disk, this value can be as high as $10^{-1}$ \citep{martin2019}, whereas for dead zones with a negligible ionization fraction, the $\alpha$ factor becomes lower than $10^{-4}$ \citep{zhu2016,martin2019}. From observations of the edge-on Oph 163131 PPD, \cite{villenave2022} deduced an $\alpha$ factor smaller than $10^{-5}$. Therefore one can extrapolate what is observed in PPD to CPDs, even if we do not have direct observations of CPD $\alpha$ factor. Moreover, this variability underscores the complex nature of disk viscosity and highlights the importance of further observational and theoretical investigations to unravel its underlying mechanisms. Consequently no constrains exist on the $\alpha$--factor of the jovian CPD.

\begin{figure*}
    \centering
    \includegraphics[width=\linewidth]{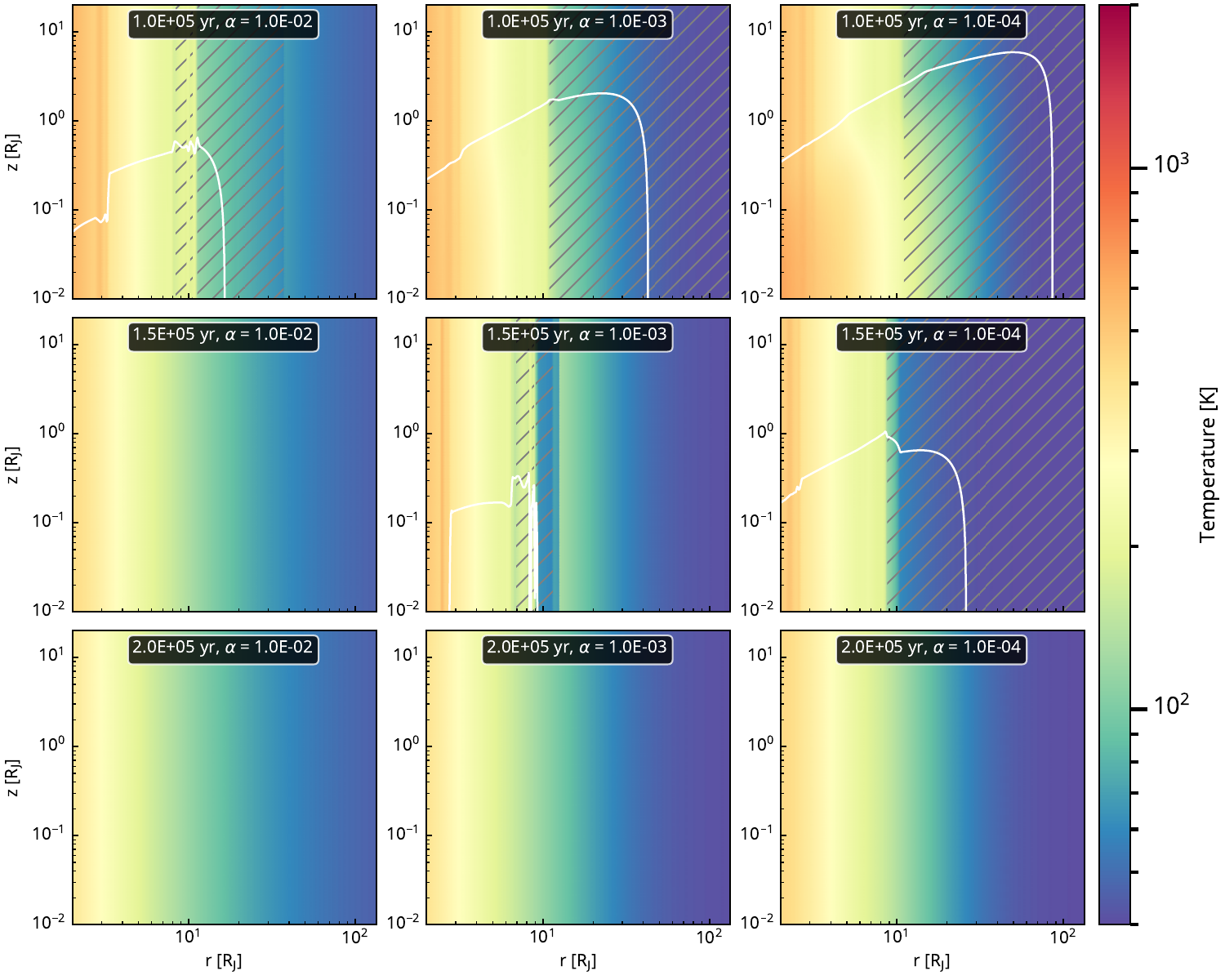}
    \caption{CPD temperature profiles with $\alpha$-viscosity parameters (from left to right columns) corresponding to $10^{-2}$, $10^{-3}$, and $10^{-4}$, depicted at $t$ = 100, 150, and 200 kyr of the CPD evolution (from top to bottom rows). The solid white line represents the altitude of the photosurface. If the photosurface is not explicitly shown or represented in a given panel, it indicates its absence. The hatched areas are located within the shadow cast by the CPD.}
    \label{fig:alpha_vs_time}
\end{figure*}

\begin{figure*}
    \centering
    \includegraphics[width=\linewidth]{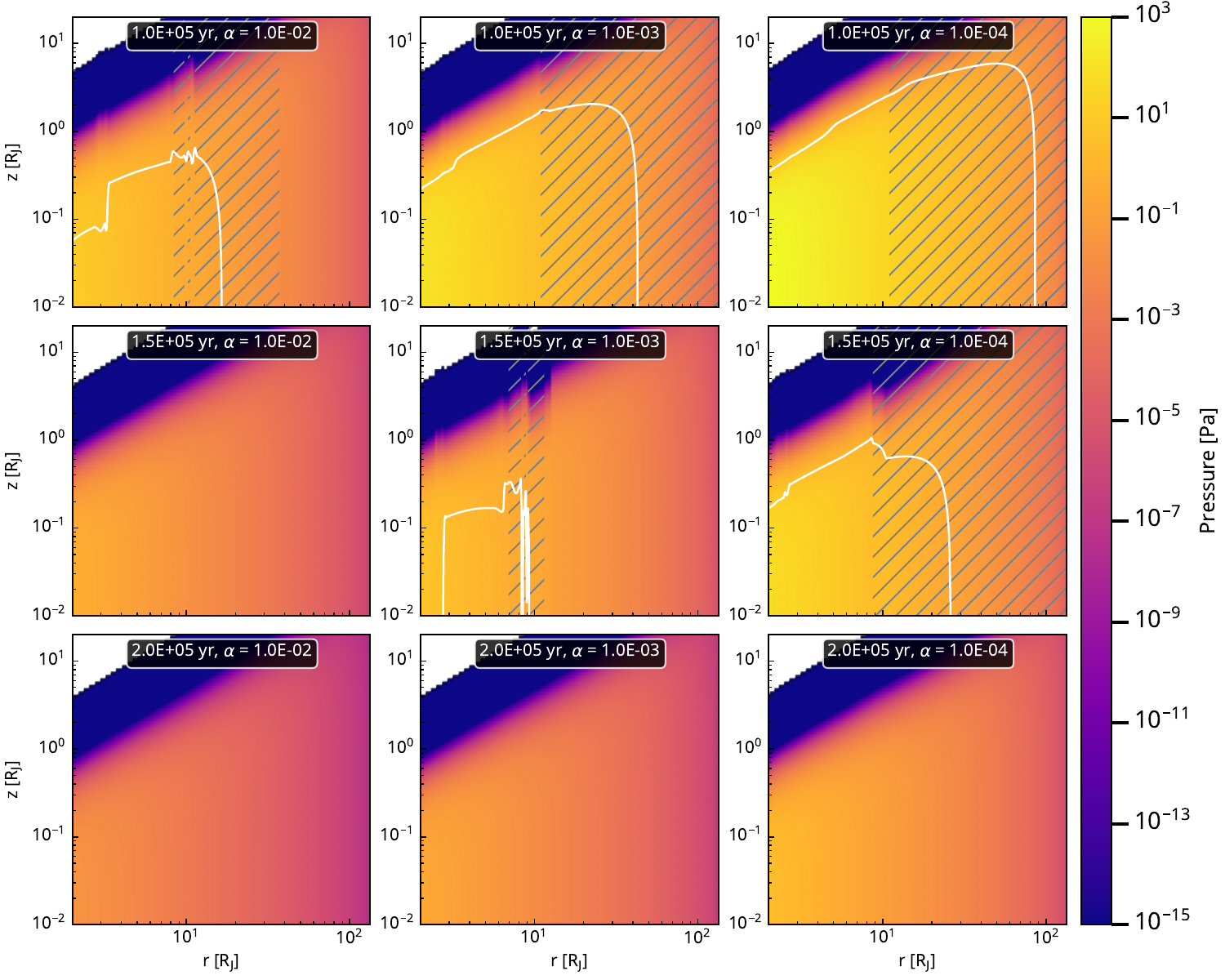}
    \caption{CPD pressure profiles with $\alpha$-viscosity parameters (from left to right columns) corresponding to $10^{-2}$, $10^{-3}$, and $10^{-4}$, depicted at $t$ = 100, 150, and 200 kyr of the CPD evolution (from top to bottom rows). The solid white line represents the altitude of the photosurface. If the photosurface is not explicitly shown or represented in a given panel, it indicates its absence. The hatched areas are located within the shadow cast by the CPD.}
    \label{fig:alpha_vs_time_pressure}
\end{figure*}

To assess the robustness of our model, we explored a range of $\alpha$ values spanning from $10^{-4}$ to $10^{-2}$. We deliberately avoided exceeding $10^{-1}$, recognizing that gas needs to reach temperatures on the order of $10^4$ K for full ionization. Figures \ref{fig:alpha_vs_time} and \ref{fig:alpha_vs_time_pressure}  show the CPD temperature and pressure , respectively, profiles with evolution parameters presented in table \ref{table:settings}   with the exception of the $\alpha$-parameters which values are  $10^{-4}$ , $10^{-3}$ and $10^{-2}$.

As we adjusted the $\alpha$ factor from $10^{-3}$ to $10^{-4}$, we observed a trend where the disk grew hotter and denser, whereas shifting the $\alpha$ factor from $10^{-3}$ to $10^{-2}$ resulted in a lighter and cooler disk. The mass of the disk exhibited a pronounced dependency on the $\alpha$ factor. For instance, with an $\alpha$ factor of $10^{-4}$, the CPD achieved a maximum mass of $7 \times 10^{-4}$ \mj, whereas with an $\alpha$ factor of $10^{-2}$, it only reached a maximum mass of $3 \times 10^{-5}$ \mj. The strong dependence of the CPD mass on the $\alpha$ parameter is a direct consequence of the effect of viscosity on the gas diffusion rate. As the $\alpha$ value increases, the viscosity increases and the viscous diffusion timescale decreases. The shorter the timescale, the faster the gas will fall onto the planet or diffuse out of the Hill sphere, resulting in a lighter and colder CPD. This is illustrated by Eq. \ref{eq:surface_density}, which shows that the disk surface density is inversely proportional to the viscosity. 

This drastic dependence on the magnitude of the $\alpha$ factor significantly influences the behavior of the photosurface altitude and the casting of shadows onto the CPD. A smaller $\alpha$ factor results in a denser CPD, thereby elevating the photosurface altitude. Consequently, with an $\alpha$ factor of $10^{-4}$, CPD regions beyond 9 \rj~remain shadowed from 130 kyr to 197 kyr after CPD's formation. Increasing the $\alpha$ factor to $10^{-2}$ leads to shadowing of regions beyond 9 \rj~occurring from 94 kyr to 120 kyr after the formation of the CPD. In other words, a smaller alpha allows the survival of shadowed regions over longer timescales. 

\section{Discussion}
\label{sec:discussion}

In this study, we demonstrate that, employing an accretion rate prescription from \cite{mordasini2013}, the CPD becomes fully vertically isothermal 161 kyr after its formation with a metallicity that is 0.1 times the protosolar value, and after 230 kyr with a 10 times protosolar metallicity. Such shadowed areas exhibit temperature reductions of approximately 100 K compared to the surrounding gas. In colder regions, the CPD's density and pressure exceed those of the surrounding environment. We illustrate that this phenomenon is highly contingent on the CPD's gas metallicity, which we vary between 0.1 and 10 times the protosolar metallicity. 

In the presented simulations, we employ the accretion rate derived from \cite{mordasini2013}, which rapidly decreases following the runaway gas accretion phase. This accretion rate is based on a model of Jupiter's formation that takes into account both the formation of a gap in the protoplanetary disk \citep{dangelo2010a} and the accretion limited by the escape of hot gas from the Hill sphere \citep{mordasini2012a}.

\subsection{Accretion rate prescription}
\label{sec:sasaki_accretion}

The accretion rate decay timescale used in our model is significantly shorter than the one typically used in most CPD models \citep{lunine1982,canup2002,mousis2002,mousis2004,sasaki2010,anderson2021,mousis2023}. Several alternative models \citep{anderson2021,mousis2023} are based on the hypothesis presented in \cite{sasaki2010} that the CPD dissipation timescale is equivalent to the protoplanetary disk dissipation timescale. To account for this major difference, we conducted tests using a slower accretion decay rate, specifically employing a CPD depletion timescale of 1 Myr as proposed by \cite{sasaki2010}. The time evolution of the accretion rate is given by:

\begin{equation}
    \dot{M}(t) = \dot{M}_0 \exp{\left(\frac{-t}{10^6 \mathrm{yr}}\right)}
    \label{eq:sasaki_accretion}
\end{equation}
With $\dot{M}_0=5\times 10^{-7}$ \mj ~yr$^{-1}$  the accretion rate at initialization. This accretion rate choice is based on the observed accretion rates, from hydrogen Balmer-line emission of the accreting Jupiter-sized exoplanets PDS~70 b and c with estimated accretion rates in the range between $10^{-8}$ and $10^{-7}$ \mj~yr$^{-1}$ \citep{aoyama2019,christiaens2019,benisty2021}. 

\begin{figure*}
\centering
\includegraphics[width=\linewidth]{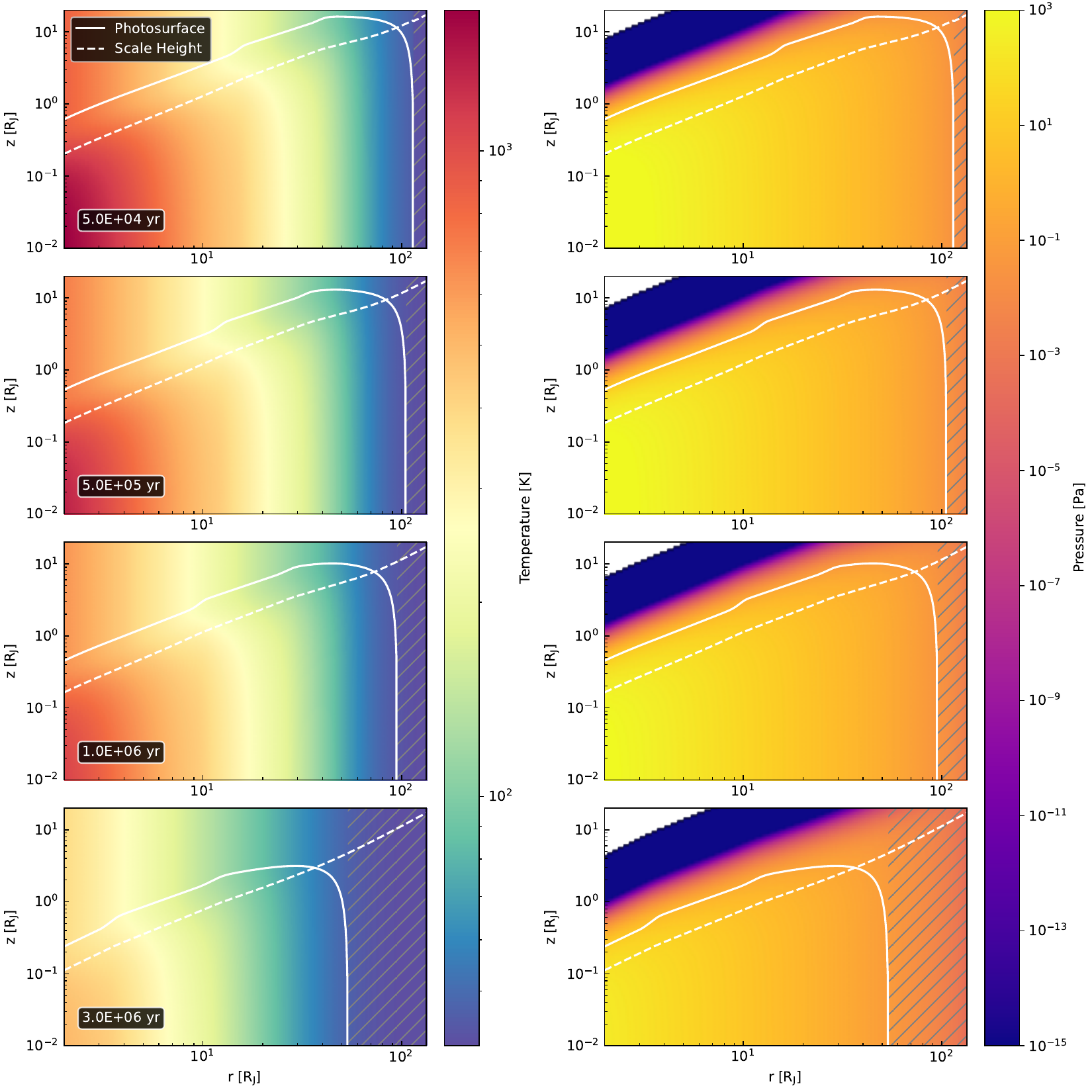}
\caption{Temperature and pressure profiles within the CPD after $t$ = 50, 500, 1000, and 3000 kyr of evolution, using the accretion rate prescription from \cite{sasaki2010}, and a metallicity 0.1 times the protosolar one. The photosurface and the CPD's scale height are represented by the white plain and dashed lines respectively.  The hatched area is located within the shadow cast by the CPD.}
\label{fig:temperature_sasaki}
\end{figure*}

Figure \ref{fig:temperature_sasaki} illustrates the temperature (left panels) and pressure (right panels) profiles at various time intervals (50, 500, 1000, and 3000 kyr) during the CPD evolution, with a metallicity  0.1 times the protosolar one. It demonstrates the slow depletion of the CPD over a period of 3000 kyr. Consequently, the CPD remains hotter after 3000 kyr of its evolution than the nominal case described in Section \ref{sec:results} after 200 kyr. This highlights the significant dependence of our results on the accretion rate evolution. Since accretion and viscous heating dominate over the planet's heating, the shadowed areas (hatched areas) do not exhibit significant temperature drops. Therefore, the consequences of CPD self-shadowing presented in Section \ref{sec:results} only exist for a rapidly decreasing accretion rate after Jupiter's formation.

One can also  vary the initial accretion rates $\dot {M}_0$. Recent three-dimensional models show that the accretion rates onto the planet can go as high as $10^{-4}$~\mj~yr$^{-1}$  at the end of its gas runnaway accretion phase \citep{lambrechts2019,schulik2020}.  Furthermore, \cite{choksi2023} conducted a survey on PPD gaps. Assuming that the observed gaps were carved by an undetected protoplanet, authors link the measured gap depth to an upper limit of the gas accretion rate onto the planet as well as its mass. \cite{choksi2023} show that the deduced accretion rates range between $8\times 10^{-10}$~\mj~yr$^{-1}$ (TWHya) and $2.5\times 10^{-5}$ (HD163296).

We increased the  initial accretion rate to  $10^{-4}$~\mj~yr$^{-1}$, which increase the CPD mass and midplane temperature up to 8000~K.  In such a case, accretion heating and viscous heating dominate the energy budget over Jupiter's radiative heating. However, at such a temperature, three--dimensional models show that we enter in the territory of the circumplanetary envelope and leave the CPD regime. Therefore even if the model converge to a profiles, since by design it compute a CPD profile, its results might not be representative of the physical reality.

\subsection{Mass of solids}
\label{sec:solid_mass}

Our nominal model considers a rapid depletion of the CPD, based on the prescription of \cite{mordasini2013}. We start the CPD evolution when Jupiter has reached 95\% of its mass. Under these conditions, the disk mass is at most $10^{-4}$ \mj, which is about half the total mass of the Galilean satellite system. Therefore, regardless of the specific dust-to-gas ratio we consider, at any given time there is not enough solid material available to form the entire Galilean satellite system instantaneously. 

However, the cumulative mass of accreted material, including both gas and dust, throughout a simulation period of 400 kyr, amounts to $1.5 \times 10^{-2}$ \mj, a much higher value than the total mass of the four Galilean moons ($\sim$$2.07 \times10^{-4}$ \mj).  Consequently, if it is assumed that the solid material required for the formation of the moons originates from the accreted material, a dust-to-gas ratio exceeding  $1.4 \times 10^{-2}$  is necessary for the accumulation of sufficient solid material onto the CPD in order to enable the formation of the Galilean satellites, based on the premise of a short-lived CPD.  Under this assumption, the accreted gas would present at minimum a protosolar metallicity. However, this finding contradicts outcomes from three-dimensional hydrodynamic simulations that trace the destiny of dust and grains in the vicinity of Jupiter's formation. These models reveal a significant depletion of grain and dust material, reduced by a factor of 100 compared to the PSN \citep{lambrechts2012,zhu2012}. Implementing this depletion factor into our model suggests that only $1.4 \times 10^{-4}$ \mj~of dust and grains would be assimilated by the CPD, over the course of the simulation. It would then prevent the formation of the moons over a decent timescale. In contrast, alternative three-dimensional hydrodynamic simulations suggest that meridional circulation effectively transports dust from the midplane of the PSN to the CPD \citep{szulagyi2022}. This mechanism results in the accreted material displaying a dust-to-gas ratio exceeding that of the PSN. In light of these conditions, there would be ample material available for the formation of the Galilean Satellites.

\subsection{Heating by cosmic ray and solar irradiation}

Our model does not account for potential gas heating caused by cosmic rays and high-energy solar irradiation emitted by the young Sun during its T-Tauri phase, particularly from solar flares. However, this heating could be significant in the optically thin regions of the CPD, especially within shadowed areas.

Cosmic and X-ray heating occurs through exothermic chemical reactions induced by radiation. The magnitude of this heating depends on the CPD's composition, particularly its molecular and electron fractions \citep{glassgold2012a}. Since the current model does not calculate these parameters, the extent of this heating cannot be directly estimated. However, \cite{glassgold2012a} suggests that such heating could contribute up to 50\% of the heat budget in molecular clouds.

Another potential source of heating from high-energy radiation is turbulence driven by magnetorotational instability (MRI) \citep{balbus1991}. This instability arises from the ionization of CPD gas, which is not evaluated in our model. Similar to radiation-induced chemical heating, MRI-driven heating depends on the CPD's composition, particularly on charge carriers such as electrons, ions, and polycyclic aromatic hydrocarbons \citep{thi2019} and references therein.

\subsection{Consequences for the formation of the Galilean moons}
\label{sec:icelines}

In our investigation, we introduced the self-shadowing of the CPD by its opaque photosurface. We illustrate that these shadows emerge in the outskirt of the disk and induce a notable temperature decline, particularly when Jupiter's surface luminosity supersedes viscous and accretion heating as the primary heat source. These shadows persist for durations ranging from 10 to 100 kyr, with their longevity directly correlated to the disk's metallicity-—the higher the metallicity, the longer the shadows endure. Within these shadowed zones, local pressures elevate by a factor of up to 2, reaching midplane pressures on the order of $10^{-1}$ Pa after 100 kyr  of simulation. Concurrently, alongside the temperature decrease and pressure increase, midplane density and surface density experience a tenfold augmentation, compared to surrounding regions.

The potential impact of high-pressure, high-density regions on the processes related to moon formation around giant planets remains uncertain. On one hand, these regions could act as dust traps, fostering areas with significantly heightened dust-to-gas ratios where proto-moons could form via streaming instability \citep{shibaike2017,shibaike2019,zhangZhang2021,chen2020}. On the other hand, shadows exert a significant influence when the CPD mass is less than $10^{-5}$ \mj, extending down to $10^{-8}$ \mj. During this evolutionary stage, the moon formation process may have transpired much earlier, possibly within the initial kyr of CPD evolution \citep{drazkowska2018}. Alternatively, moon formation could even occur during the runaway accretion phase itself, within a much more massive circumplanetary envelope, with a mass up to 0.5 \mj~\citep{keith2014}

Furthermore, the  launch of the James Webb Space Telescope (JWST) has enabled extensive observations and studies of the surface composition of the Galilean moons. The most notable observation so far is the putative detection of CO$_2$ ice on Europa's surface, which is believed to originate from the internal oceans of the moons \citep{trumbo2023}.

\begin{figure*}[h!] 
\centering
\includegraphics[width=\linewidth]{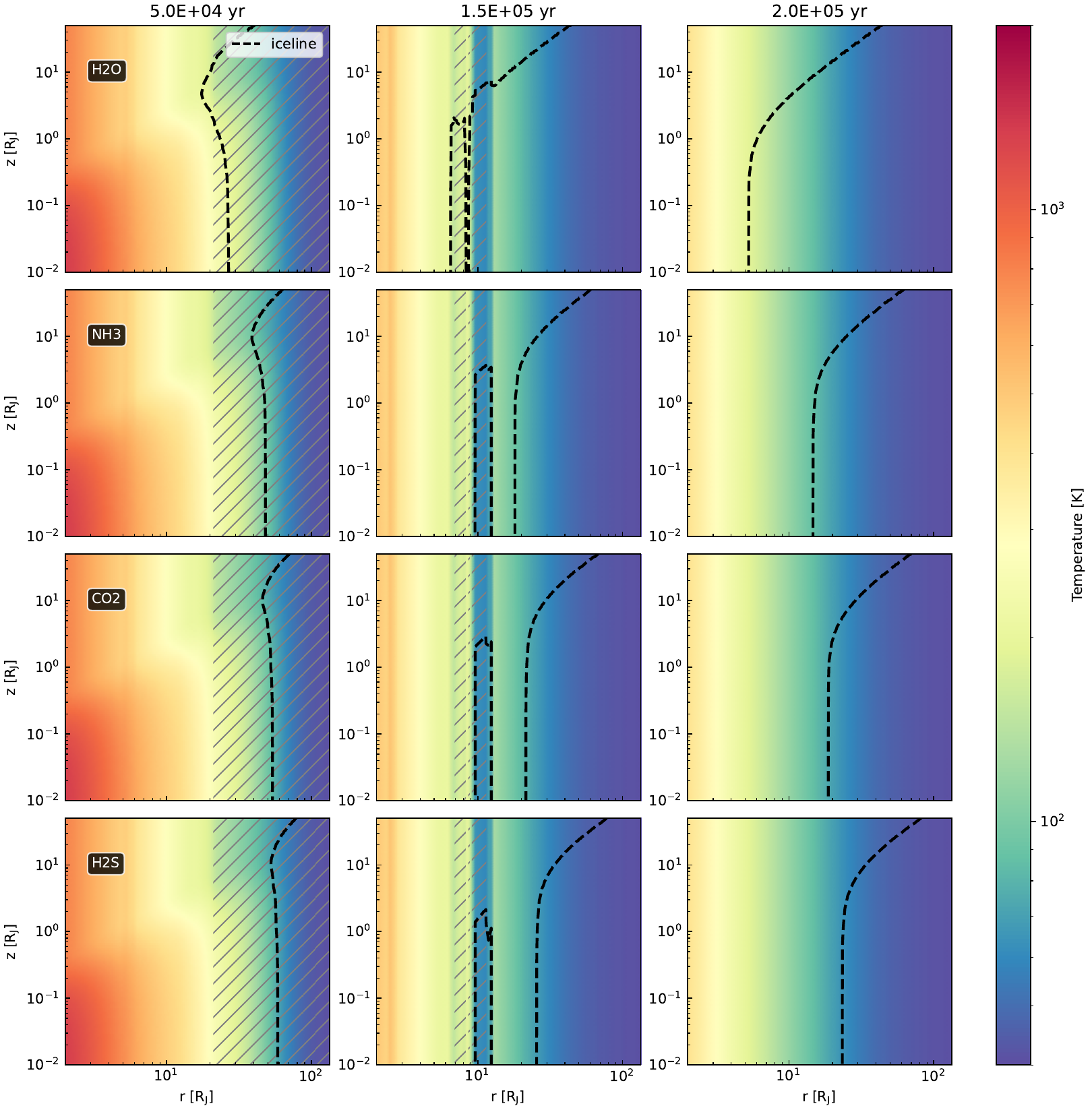}
\caption{From top to bottom: shapes of the H$_2$O, NH$_3$, CO$_2$ and H$_2$S icelines (black dashed lines) at $t$ = 50, 150 and 200 kyr of the CPD evolution.  The hatched areas are located within the shadow cast by the CPD. We use the nominal model depicted in Sec. \ref{sec:results}, with a metallicity  0.1 times protosolar.}
\label{fig:snowline}
\end{figure*}

Our model can be used to trace the evolution of volatile species within the CPD. We examine the positions of the condensation lines for a large set of volatile species H$_2$O, CO, CO$_2$, CH$_4$, H$_2$S, N$_2$, NH$_3$, Ar, Kr, Xe, and PH$_3$). Out of the eleven volatile species examined, only four (H$_2$O, NH$_3$, CO$_2$, and H$_2$S) have condensation temperatures higher than the minimum temperature of 40 K in our CPD model. Figure~\ref{fig:snowline} shows the positions of the condensation lines for these four volatile species at 40, 150, and 200 kyr of the CPD evolution. The black dashed lines in the figure represent the condensation lines for the various volatile species within the CPD. At 40 kyr after the formation of the CPD, the disk is still hot and massive, leading to the condensation lines being located far from Jupiter. In the disk midplane, the condensation lines are positioned at 34.5 \rj, 58.2 \rj, 64.4 \rj, and 70.2 \rj~for H$_2$O, NH$_3$, CO$_2$, and H$_2$S, respectively. In contrast, at 200 kyr, when the CPD has become much colder, the condensation lines have moved significantly inward, to 5.2 \rj, 14.6 \rj, 18.6 \rj, and 23.3 \rj~for H$_2$O, NH$_3$, CO$_2$, and H$_2$S, respectively. Between these two extremes, disk self-shadowing comes into play, and produces a cold area centered around 10 \rj. At 150 kyr of CPD evolution, this shadowed region acts as a cold trap for the volatile species.
Within the 9 \rj~to 12 \rj~range, all four volatile species can form ices at distances closer than their respective condensation lines, which are located at 18 \rj, 21.5 \rj, and 25.5 \rj~for NH$_3$, CO$_2$, and H$_2$S, respectively. This cold trap exists at epochs between 127 and 160 kyr after the CPD formation, and could have consequences on the composition of the grains within the disk.

Furthermore, the presence of a shadowed regions in the CPD could induce a local enhancement of the gas-phase C/O and N/O ratios as carbon and nitrogen bearing species condenses,  as shown by \cite{ohno2021a} and \cite{notsu2022}. Authors show that in the vicinity of the condensation lines the C/O ratio can reach over unity and N/O ratio be larger than 10. Such a compositional signature could be used in future CPD observation around exoplanets to assess the existence of such cold traps.

The analysis of the volatile condensation lines within the CPD suggests that disk self-shadowing could play a key role in understanding the early icy grain composition and, consequently, the composition of the building blocks of the Galilean moons. While this analysis is qualitative rather than quantitative, it demonstrates that the cold trap created by the self-shadowed region of the disk could lead to the formation of ices for various volatile species, such as H$_2$O, NH$_3$, CO$_2$, and H$_2$S, at distances closer to Jupiter than their respective condensation lines. This effect on the icy grain composition within the disk could be significant, as the volatile species would be able to condense and accrete onto the grains at closer distances to the planet. This, in turn, could influence the overall composition of the material that eventually forms the Galilean moons. 

However, this approximation has  limitations. First, it depend on the timing of formation of the moons, which building blocks have to form before 1 Myr of  CPD evolution, otherwise, the shadows' influence are weaker on the thermal and pressure profiles of the CPD. Therefore, if Ganymede and Callisto formed at times beyond 1.5 Myr after Jupiter's formation as suggested by \cite{shibaike2017,shibaike2019}, cold traps would have a limited impact on the moons composition. Furthermore, Ganymede and Callisto contains much ices than Europa, \citep{kuskov2005} , probably showing that the two moons formed in a colder environment. However, the volatile content on the formed moon will depend on the timing formation. In fact if a moon accreted early in solar system history, its interior will be heated radiogenic heating induced by short lived isotopes such as \ce{^{26}Al}  \citep{monnereau2013}. Consequently, Europa could have formed early in CPD evolution, typically before 1 Myr  of evolution, and lost a part of its hydrosphere because radiogenic heating. 

\section{Summary and conclusion}
\label{sec:summary}

This study demonstrates the significant impact of Jupiter's heating and the resulting disk self-shadowing on the evolution of the thermal structure within a rapidly depleting CPD.

Our key findings are the following: 

\begin{enumerate}
\item {\it Metallicity Dependence.} Our results show a strong dependency on the metallicity of the accreted material, with longer-lasting shadows in the CPD evolution for more metallic disks.

\item {\it Parameter Variability.} We varied the model parameters to assess the variability of our results, which confirmed the robustness of the self-shadowing effect.

\item {\it Accretion Rate Dependence.} The self-shadowing effect is also dependent on the decay timescale of the accretion rate. The longer an accretion rate lasts, the less significant the impact of the shadow on the thermal structure of the CPD.

\item {\it Cold Traps for Volatile Ices.} The shadows cast onto the disk could have produced cold traps where volatile ices, such as NH$_3$, CO$_2$, and H$_2$S, were trapped at epochs between 127 and 160~kyr, and therefore for a duration period of 35~kyr. With increasing CPD metallicity, this duration periods can be extended up to 120 kyr at distances centered around 10 \rj.

\end{enumerate}

The implications of the self-shadowing effect and resulting cold traps for volatile ices within the CPD could be the following for the composition of the Galilean moons:

\begin{enumerate}

\item The cold traps centered around 10 \rj, where volatile species like NH$_3$, CO$_2$, and H$_2$S could condense at closer distances to Jupiter, may have influenced the overall composition of the building blocks that eventually formed the Galilean moons.

\item If the cold traps had a significant impact, it is possible that the inner Galilean moons, such as Europa, could contain a greater proportion of these volatile species compared to the outer moons like Callisto.

\end{enumerate}

Our hypothesis regarding potential compositional differences between the inner and outer Galilean moons due to the self-shadowing effect can be further investigated and tested by the upcoming ESA/JUICE and NASA/Europa-Clipper missions, which will provide detailed observations of these icy worlds.

\begin{acknowledgements}
The project leading to this publication has received funding from the Excellence Initiative of Aix-Marseille Universit\'e--A*Midex, a French ``Investissements d’Avenir program'' AMX-21-IET-018. This research holds as part of the project FACOM (ANR-22-CE49-0005-01\_ACT) and has benefited from a funding provided by l'Agence Nationale de la Recherche (ANR) under the Generic Call for Proposals 2022. OM acknowledges funding from CNES.
\end{acknowledgements}

\bibliography{CPD}

\end{document}